\documentclass[fleqn,twoside]{article}
\usepackage{espcrc2}
\usepackage{epsfig}
\usepackage{graphicx}
\usepackage[figuresright]{rotating}
\newcommand{\bc} {\begin{center}}
\newcommand{\ec} {\end{center}}
\newcommand{\bqa}{\begin{eqnarray}}
\newcommand{\eqa}{\end{eqnarray}}
\newcommand{\nn}{\nonumber}
\newcommand{\al}{\alpha}
\newcommand{\be}{\beta}
\newcommand{\ga}{\gamma}
\newcommand{\de}{\delta}
\newcommand{\ep}{\epsilon}
\newcommand{\ph}{\phi}

\newcommand{\ta}{\tau}
\newcommand{\ka}{\kappa}
\title{\vskip -110pt
\mbox{} \hfill BI-TP 2002/20\\
\mbox{} \hfill DESY 02-117\\
\mbox{} \hfill August 2002\\
\vskip 55pt
The H dibaryon on the lattice}

\author{I. Wetzorke\address{NIC / DESY Zeuthen,
Platanenallee 6, 15738 Zeuthen, Germany}
and F. Karsch\address{Fakult\"at f\"ur Physik,
Universit\"at Bielefeld, 33615 Bielefeld, Germany}}
       
\begin{document}

\begin{abstract}
We present our final results for the mass of the six quark flavor
singlet state ($J^P=0^+$, $S=-2$) called H dibaryon, which would be the
lightest possible strangelet in the context of strange quark matter.
The calculations are performed in quenched QCD on $(8-24)^3 \times 30$
lattices with the (1,2) Symanzik improved gauge action and the clover
fermion action. Furthermore the fuzzing technique for the fermion fields
and smearing of the gauge fields is applied in order to enhance the
overlap with the ground state. Depending on the lattice size we observe
an H mass slightly above or comparable with the $\Lambda\Lambda$
threshold for strong decay. Therefore a bound H dibaryon state seemed
to be ruled out by our simulation.
\vspace{1pc}
\end{abstract}

\maketitle

\section{INTRODUCTION}
A bound six quark state ($uuddss$) called H dibaryon with a mass
81 MeV below the 2231 MeV $\Lambda\Lambda$ threshold for strong decay
was predicted in a bag model calculation in 1977 \cite{Jaf77}.
The H dibaryon is the lightest possible SU(3) flavor
singlet state with spin zero, strangeness \mbox{-2} and $J^P=0^+$.
It has been speculated that the H dibaryon could form a Bose condensate in
nuclear matter prior to the quark-hadron phase transition of QCD at
high density. In the context of strange quark matter the H dibaryon
was suggested as a candidate for the smallest strangelet
\cite{Sch98}, which became relevant in the discussion
on disaster scenarios at RHIC \cite{Bus00}.

In the 25 years after the initial prediction of a stable H dibaryon
state numerous searches for this particle were performed. On the
theoretical side this involved various model calculations, QCD sum
rules and perturbative calculations including color-spin (One-Gluon-Exchange)
or flavor-spin (Instanton Induced Interaction, Goldstone-Boson-Exchange)
dependent q-q interactions. The resulting H dibaryon masses scatter
in a range of ${\cal O}$(100 MeV) around the $\Lambda\Lambda$ threshold.
Moreover experimental searches showed almost no evidence for a
bound H dibaryon. The review article \cite{Sak00} summarizes both
the theoretical and experimental attempts to verify the existence and stability
of the H dibaryon.

In the 1980's two lattice calculations \cite{Mac85,Iwa88} reported
contradictory results on $m_H$, but both simulations suffered from heavy quark
masses and low statistics. In 1999 yet another calculation \cite{Neg99}
overcame these problems and concluded that the H dibaryon state is unbound in
the infinite volume limit. Our aim was to check these results with refined
methods, which includes on the one hand the use of improved gauge and fermion
actions and on the other hand the use of smearing and fuzzing techniques
to enhance the overlap with the ground state.

\section{DETAILS OF THE SIMULATION}\label{sim}
We performed our calculations with the (1,2) Symanzik improved gauge
action and the tree-level improved clover fermion action on rather coarse
lattices ($\be=4.1$, $a=0.177(8)$ fm) of the size $(8-24)^3\times 30$ in order
to study the large six quark object with an adequate spatial lattice extent.
Depending on the lattice size we generated 40 to 120 configurations
on which the propagators for three to five kappa values for the degenerate
$u/d$ and heavier $s$ quark masses in the range of 30 to 250 MeV were
calculated.

The particle masses are obtained from exponential decrease of the
correlation function $G(\ta) = \int d^3 x \langle\; {\cal O}(\vec{x},\ta)
{\cal O}^\dag(\vec{0},0)\;\rangle$ for the appropriate operators of the
$\Lambda$ baryon and the H dibaryon \cite{Gol92}
(color indices: roman letters, spinor indices: greek letters):
\bqa
{\cal O}_\Lambda(x)\!\!\!&=&\!\!\!\ep_{abc} (C\ga_5)_{\be\ga}
[ u_{\al}^a(x) d_{\be}^b(x) s_{\ga}^c(x) \nn\\
&&\hspace*{-3.5mm}+d_{\al}^a(x) s_{\be}^b(x) u_{\ga}^c(x)
\!\!-\!2 s_{\al}^a(x) u_{\be}^b(x) d_{\ga}^c(x) ] \nn\\
{\cal O}_H(x)\!\!\!&=&\!\!\!3 (udsuds) -3 (ussudd) -3 (dssduu) \nn\\
({\sf abcdef})\!\!\!\!\!&=&\!\!\!\!\!\ep_{abc} \ep_{def}
(C\ga_5)_{\al\be} (C\ga_5)_{\ga\de} (C\ga_5)_{\ep\ph} \nn\\
&&\;*\;{\sf a}_{\al}^a(x) {\sf b}_{\be}^b(x) {\sf c}_{\ep}^c(x)
{\sf d}_{\ga}^d(x) {\sf e}_{\de}^e(x) {\sf f}_{\ph}^f(x) \nn
\eqa
In addition to the $\Lambda$ and H dibaryon we calculated the
correlators of the (non-)strange mesons $\pi, \rho, K, K^*$ as
well as the baryons $N$ and $\Sigma$. These particle masses were
used for scale setting and comparison to the experimental spectrum.
The degenerate $u/d$-quark mass is fixed by the ratios
$m_\pi/m_N$, $m_\pi/\sqrt\sigma$ and $m_N/\sqrt\sigma$ yielding
a common value of $\ka_{ud}$=0.1490(1). In order to eliminate the
quenching effects most efficiently in the setting of the strange
quark mass, we determined a mean value of $\ka_s$=0.1417(2) by
averaging over the $\ka_s$-values obtained from the particle ratios
$m_\Lambda/m_N$, $m_\Sigma/m_N$ and $m_{K^*}/m_N$. Furthermore the
dependence of the mass difference $m_H - 2 m_\Lambda$ on the three
possible ways of fixing $\ka_s$ is investigated in section \ref{res}.

\subsection{Smearing \& Fuzzing}
A good projection to the ground state is very important for a reliable
extraction of the particle masses. A better overlap with the ground
state can be achieved by the iterative smearing of the gauge links and
the application of the fuzzing technique for the fermion fields \cite{Lac95}
taking the physical size of the particle into account. The separation of a
quark-quark pair by a suitable distance, the so-called fuzzing radius, thus
will maximize the ground state contribution relative to the ones of the excited
states already at small Euclidean time separations. We apply this method for
the calculation of the meson, baryon and dibaryon correlators.  
We could show \cite{Wet00} that the largest plateau in the effective mass
is obtained for four fuzzed quarks, actually the four light quarks inside
the H dibaryon, employing a fuzzing radius of about 0.7 fm. For the strange
mesons (baryons) we use the one (two) light quark field(s) fuzzed with the
same fuzzing radius.

\section{RESULTS}\label{res}
\begin{figure}[t]
\bc
\hskip-4mm
\epsfig{file=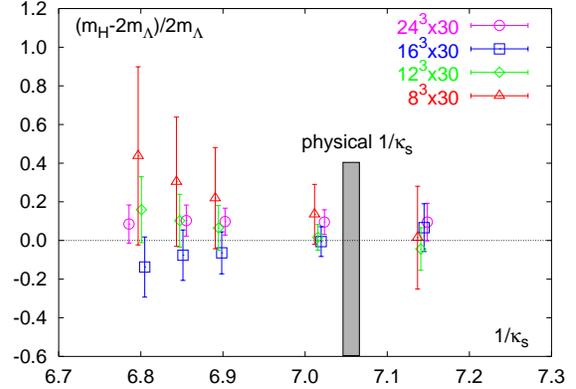,width=78mm}
\vskip-12mm
\ec
\caption{Dependence of the mass difference on the
lattice size and $\ka_s$ at physical $\ka_{ud}$}
\label{ks}
\end{figure}

The particle masses calculated for different kappa values are first
extrapolated linearly to the physical $\ka_{ud}$ value while keeping $\ka_s$
fixed. The final result is determined by interpolating to the mean $\ka_s$
value, where the physical scale is fixed by the particle ratios as mentioned
above. The obtained $K^*$, $\Sigma$ and $\Lambda$ masses are compared to the
experimental values and show a deviation less than 8 \% except for the smallest
lattice size. Such a variance can be considered as usual quenching effect.
Relatively strong finite size effects are observed for the smallest lattice
size $L = 8 a = 1.4$ fm. Whereas the strange mesons masses obtained on this
lattice are only slightly higher than the ones on the larger lattices, this
effect is more pronounced for the heavier baryon masses. For the six quark H
dibaryon state a very strong influence of the spatial lattice size is
observed. In figure \ref{ks} we show the mass difference
$m_H - 2 m_\Lambda$ at physical $\ka_{ud}$ for the four different lattice
sizes. The grey band indicates the mean $\ka_s$ value as explained in section
\ref{sim}. The points are slightly displaced in $\ka_s$ for better visibility.
The slope of the $\ka_s$ dependence is obviously quite different for varying
lattice volume, but the mass difference in the physical region is nevertheless
quite similar and near zero for all investigated lattice sizes.

\begin{figure}[t]
\bc
\hskip-4mm
\epsfig{file=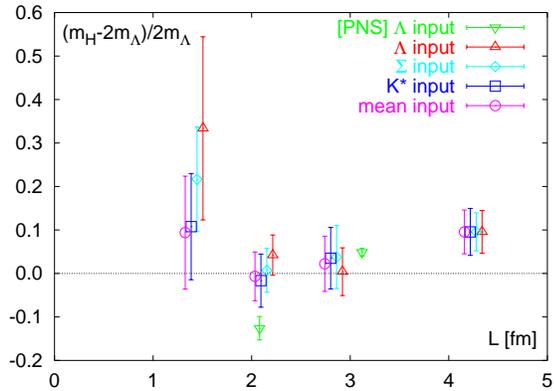,width=78mm}
\vskip-14mm
\ec
\caption{Dependence of the mass difference on the choice of
scale setting particle at physical $\ka_{ud}$ and $\ka_s$
([PNS] values for $\Lambda$ input values are taken from \cite{Neg99}.)}
\label{phys}
\end{figure}

In order to clarify the evidence for an unbound H dibaryon state, a closer look
at the dependence of the scale setting by the particle ratios seems helpful.
Figure \ref{phys} shows the mass difference for all lattice sizes using the
$\Lambda$, $\Sigma$ or $K^*$ masses as input to fix $\ka_s$ as mentioned above.
The points are again a bit displaced - this time in the spatial lattice extent
$L$. Moreover the interpolation to the mean value of $\ka_s=0.1417(2)$ and the
previous results of Pochinsky et al.~\cite{Neg99} are shown to allow a direct
comparison. The strong finite size effects are probably the reason for the
overestimated mass difference observed for the smallest lattice size.
On the intermediate lattice sizes all values of $m_H - 2 m_\Lambda$ are
compatible with zero independent of the particle ratio used as input. On the
largest  lattice with $L/a=24$ the values are slightly higher, but may come
closer to zero with larger statistics. Therefore our simulations provide no
evidence of a bound H dibaryon state for the investigated lattice sizes. A
detailed table of the particle masses for the respective choice of input
particle will be given soon in a longer write-up \cite{Kar02}.

Finally we want to compare our findings with the previous lattice results
\cite{Neg99}, which were also obtained on $L/a=16$ and 24 lattices but with a
smaller lattice spacing of 0.13(3) fm. At a spatial size of about 2 fm the
former result lies at bit below our values, while a good agreement can be
observed for a larger extent of $L\approx 3$ fm. Hence a common conclusion
arises from the recent and present studies: The H dibaryon does not exist as
stable particle in the vacuum, at least in the limit of quenched QCD. 
\section{SUMMARY \& OUTLOOK}
We have presented the results of the first lattice investigation on the
H dibaryon state employing improved gauge and fermion actions, relatively
light quark masses as well as smearing and fuzzing techniques to enhance
the overlap with the ground state of the particle. We observe a H dibaryon
mass slightly above or comparable with the $\Lambda\Lambda$ threshold
for strong decay on all lattice sizes. We thus provide further evidence 
for an unbound H dibaryon state consisting of two lambda baryons.

A final remark concerns the relative error of the H dibaryon correlation
function, which is rising only linearly compared to the $\Lambda$ baryon
correlator. This observation raises the hope that simulations of larger
strangelets or even multi-quark clusters might be possible in the future.

\end{document}